\documentclass[14pt]{article}
\usepackage[a4paper, margin=1in]{geometry}
\usepackage{graphics}
\usepackage[absolute,overlay]{textpos}
\usepackage{amsmath}
\usepackage{xspace}
\usepackage{graphicx}
\usepackage{comment}
\usepackage{lineno}
\usepackage{url}
\usepackage{hyperref}
\usepackage{caption}
\newcommand{\phis}{\ensuremath{\phi_s}\xspace}
\newcommand{\BsJPsiPhi}{\ensuremath{B_{s}\rightarrow J/\psi \phi}\xspace}

\newcommand{\Bs}{\ensuremath{B_{s}}\xspace}

\def\jpsiml{\ensuremath{3.07}}
\def\jpsimr{\ensuremath{3.14}}
\def\phiml{\ensuremath{1.017}}
\def\phimr{\ensuremath{1.023}}
\def\bsml{\ensuremath{5.28}}
\def\bsmr{\ensuremath{5.46}}

\def\lhcbNBs{\ensuremath{117000}}
\def\lhcbRes{\ensuremath{0.041~\mathrm{rad}}}
\def\lhcbxi{\ensuremath{0.0186}}
\def\lhcbtime{\ensuremath{45.5~\mathrm{fs}}}
\def\lhcbtag{\ensuremath{4.73\%}}
\def\lhcbratio{\ensuremath{2.28}~\mathrm{rad}}
\begin{document}
% \linenumbers
%
\title{Prospect for measurement of the CP-violating phase~\phis in the \BsJPsiPhi channel at a future $Z$ factory}

\author{Xiaomei Li$^1$, Manqi Ruan$^2$, Mingrui Zhao$^{1,3}$\footnote{mingrui.zhao@mail.labz0.org}
\\
\footnotesize{1.Science and Technology on Nuclear Data Laboratory, China Institute of Atomic Energy, Beijing, China}
\\
\footnotesize{2.Institute of High Energy Physics, Chinese Academy of Sciences, Beijing, China}
\\
\footnotesize{3.Niels Bohr Institute, University of Copenhagen, Copenhagen, Denmark}
}
\date{}

% \author{Xiaomei Li\inst{1}, Manqi Ruan\inst{2} \and Mingrui Zhao\inst{1,3}% etc
% \author{Xiaomei Li, Manqi Ruan and Mingrui Zhao}% etc
% \thanks{\emph{Present address:} mingrui.zhao@mail.labz0.org}%
%
% \offprints{}          % Insert a name or remove this line
%
% \institute{\inst{1}Science and Technology on Nuclear Data Laboratory, China Institute of Atomic Energy, Beijing, China \\
% \inst{2}Institute of High Energy Physics, Chinese Academy of Sciences, Beijing, China \\
% \inst{3}Niels Bohr Institute, University of Copenhagen, Copenhagen, Denmark}

%
% \date{Received: date / Revised version: date}

\begin{textblock*}{20cm}(1cm,1cm) % {block width} (x-coordinate, y-coordinate)
% This is your comment in the top-left corner
  Accepted by European Physical Journal C, the publication is available at \href{https://link.springer.com/article/10.1140/epjc/s10052-024-13217-3?utm\_source=rct\_congratemailt\&utm\_medium=email\&utm\_campaign=oa\_20240827\&utm\_content=10.1140/epjc/s10052-024-13217-3}{Link}
\end{textblock*}

\maketitle
\abstract{
The CP-violating phase \phis, the \Bs decay width ($\Gamma_s$), and the \Bs decay width difference ($\Delta\Gamma_s$) are sensitive probes to new physics and can constrain the heavy quark expansion theory. The potential for the measurement at future $Z$ factories is studied in this manuscript. It is found that operating at Tera-$Z$ mode, the expected precision can reach: $\sigma(\phis) = 4.6~\mathrm{mrad}$, $\sigma(\Delta\Gamma_s) = 2.4~\mathrm{ns^{-1}}$ and $\sigma(\Gamma_s) = 0.72~\mathrm{ns^{-1}}$. The precision of $\phis$ is 40\% larger than the expected precision with the LHCb experiment at HL-LHC. If operating at 10-Tera-$Z$ mode, the precision of $\phis$ can be measured at 45\% of the precision obtained from the LHCb experiment at HL-LHC. However, the measurement of $\Gamma_s$ and $\Delta\Gamma_s$ cannot benefit from the excellent time resolution and tagging power of the future $Z$-factories. Only operating at 10-Tera-$Z$ mode can the $\Gamma_s$ and $\Delta\Gamma_s$ reach an 18\% larger precision than the precision expected to be obtained from LHCb at HL-LHC. The control of penguin contamination at the future Z-factories is also discussed. 
%
% \PACS{
%       {PACS-key}{CP-Violation}   \and
%       {PACS-key}{Future colliders}
%      } % end of PACS codes
} %end of abstract
\section{Introduction}
In the Standard Model (SM), CP violation is attributed to the Cabibbo-Kobayashi-Maskawa (CKM) matrix. The CP-violating phase, denoted as $\phi_s$, emerges from the interference between the direct decay amplitude of the \Bs meson and the amplitude of the $B_s$ meson decaying after $B_s$--$\bar{B}_s$ oscillation.
In the SM,  when subleading contributions are neglected, the phase $\phis$ is predicted to be $\phi_s = -2 \beta_s$, where $\beta_s$ is defined as $\beta_s \equiv \arg{\left[-\frac{V_{ts}V^*_{tb}}{V_{cs}V^*_{cb}}\right]}$, represented by the elements of the CKM matrix. However, when accounting for the penguin diagram's contribution, the phase is modified by a shift $\Delta\phi_s$, resulting in $\phi_s = -2 \beta_s + \Delta\phi_s$.
The current SM prediction for the phase $\phi_s$ is $-0.03696^{+0.00072}_{-0.00082}~\mathrm{rad}$, according to the CKMFitter group~\cite{Charles:2015gya}, and $-0.03700\pm0.00104~\mathrm{rad}$ from the UTfit Collaboration~\cite{UTfit:2006vpt}. The global average in experiments stands at 
$\phis=-0.049 \pm 0.019~\mathrm{rad}$~\cite{Workman:2022ynf},
% $\phi_s = -0.021\pm0.031~\mathrm{rad}$
with the uncertainty being approximately 20 times larger than that of the SM prediction. Accurate measurement of $\phi_s$ serves as a critical test for the Standard Model.
 
The \Bs meson exists in two mass eigenstates, known as the light (L) and heavy (H) states, each with distinct decay widths, denoted as $\Gamma_L$ and $\Gamma_H$, respectively. Measurements of the \Bs decay width difference, $\Delta\Gamma_s \equiv \Gamma_L - \Gamma_H$, and the average decay width, $\Gamma_s \equiv (\Gamma_L + \Gamma_H)/2$, hold significant theoretical interest. The Heavy Quark Expansion (HQE)~\cite{HQE} theory provides a robust framework for calculating various observables related to $b$-hadrons. Accurate measurements of $\Gamma_s$ and $\Delta\Gamma_s$ serve as critical tests for the validity of the HQE theory.

After the Higgs boson discovery in 2012, the Circular Electron-Positron Collider (CEPC) and the Future Circular Collider (FCC-ee) were proposed. These colliders are designed not only as Higgs factories but also to operate at the $Z$ pole configuration. In this mode, they are projected to produce between $10^{12}$ and $10^{13}$ $Z$ bosons over a decade. 
Consequently, from the decay of these $Z$ bosons, approximately $0.152 \times (10^{12} - 10^{13})$ $b\bar{b}$ pairs are expected to be generated. Thus, the future $Z$-factories are proposed to serve concurrently as $b$-factories.
Using a time projection chamber or a wire chamber as the main tracking detector, the detectors at the CEPC and FCC-ee offer excellent particle identification, highly accurate track and vertex reconstruction, and extensive geometric acceptance, which are all important in heavy flavor physics study. These capabilities position the future $Z$-factories as excellent experiments for advancing heavy flavor physics research.

This paper explores the expected measurement precision at future $Z$-factories, extrapolating from measurements from current operating experiments. 
The extrapolation process is carried out as follows:
First, we list all important factors influencing measurement precision, including the statistical data size and detector performances. We then figure out the mathematical relationship determining how these factors influence measurement precision.
Subsequently, for each of these factors, we assess their performance at future $Z$-factories.
Finally, we compare the performances of these factors between the future colliders and the current existing experiments. The expected precision of the interested parameters at the future colliders is then computed by applying the mathematical relationship, using inputs from the statistical data size and detector performances.

\subsection{Measurement of \phis ($\Delta\Gamma_s$, $\Gamma_s$) in experiments}

The CP-violating phase $\phi_s$, \Bs decay width $\Gamma_s$, and the width difference $\Delta\Gamma_s$ between the heavier and lighter \Bs meson eigenstates have been thoroughly investigated in experiments conducted by ATLAS ~\cite{ATLAS:2020lbz,ATLAS:2014nmm}, CDF~\cite{CDF:2012nqr}, CMS~\cite{CMS:2015asi,CMS:2020efq}, D0~\cite{D0:2011ymu}, and LHCb~\cite{LHCb:2017hbp,LHCb:2019nin,LHCb:2014ini,LHCb:2016tuh,LHCb:2019sgv,LHCb:2023sim}. The decay channel $\Bs \rightarrow J/\psi (\rightarrow \mu^+ \mu^-) \phi (\rightarrow K^+ K^-)$ is particularly notable due to its sizeable branching fraction and the final state consisting entirely of charged tracks. This decay channel benefits from the narrow decay widths of the $J/\psi$ and $\phi$ particles, effectively suppressing the combinatorial background. It stands as the most prominent channel for measuring $\phi_s$, and it also allows for the concurrent extraction of $\Delta\Gamma_s$ and $\Gamma_s$.

The time and angular distribution of $\BsJPsiPhi$ is a sum of ten terms corresponding to the three polarization amplitudes and the non-resonant S-wave, together with their interference terms:
\begin{equation}
    \label{eq:BsJPsiPhi}
\frac{d^4\Gamma(\BsJPsiPhi)}
{dt d\Omega}
\propto
\sum_{k=1}^{10}
h_{k}(t)f_{k}(\Omega),
\end{equation} where 
\begin{displaymath}
\begin{array}{rcl}
  h_k(t|B_{s}) &=& \displaystyle N_k e^{-\Gamma_s t}
\Biggl [
a_k \cosh(\frac12\Delta\Gamma_st)  +
b_k \sinh(\frac12\Delta\Gamma_st) \\
&+& \displaystyle  c_k \cos(\Delta m_s t) + d_k \sin(\Delta m_s t) 
\Biggl ], \\
h_k(t|\bar{B}_{s}) &=& \displaystyle N_k e^{-\Gamma_s t}
\Biggl [
a_k \cosh(\frac12\Delta\Gamma_st)  +
b_k \sinh(\frac12\Delta\Gamma_st) \\
&-& \displaystyle c_k \cos(\Delta m_s t) - d_k \sin(\Delta m_s t)
\Biggl ],
\end{array}
\end{displaymath}
and $f_k(\Omega)$ is the amplitude function.

In the formulation of $h_k(t)$, the term $\Delta m_s$ represents the mass difference between the $B_s$ mass eigenstates, while $N_k$ denotes the amplitude of the component at $t=0$. The phase $\phi_s$ is encapsulated within the parameters $a_k, b_k, c_k$, and $d_k$. For an in-depth explanation of these parameters, one can refer to the LHCb publication~\cite{LHCb:2019nin}. The values of $\phi_s$, $\Delta\Gamma_s$, and $\Gamma_s$ could be obtained by fitting the time and angular distributions of $B_s \rightarrow J/\psi \phi$ decay events.

Additionally, when determining $\phi_s$, $\Delta\Gamma_s$ and $\Gamma_s$, parameters such as $\Delta m_s$ can also be simultaneously derived from this fitting. However, the precision of these parameters is beyond the scope of this work and will not be discussed here.
% [TODO] {\color{red} Additionally, together with the obtaining $\Delta\Gamma_s$, and $\Gamma_s$, the parameters such as $\Delta m_s$ could also be obtained through this fitting at the same time . Out of interests from this work, their precision will not be discussed here.}

\section{Estimation of precision on the future $Z$ factory} 
The statistical precision of the $\phi_s$ measurement, denoted by $\sigma(\phi_s)$, is directly proportional to the inverse square root of the effective signal sample size. This effective sample size is, in turn, dependent on the number of $b\bar{b}$ pairs ($N_{b\bar{b}}$) produced by the collider. Additionally, the effective signal sample size is proportional to the detector's acceptance and efficiency $\varepsilon$.

Identifying the initial flavor, either $B_s$ or $\bar{B}_s$, is essential for extracting parameters from Eq.~(\ref{eq:BsJPsiPhi}). This procedure is known as flavor tagging.
The tagging efficiency, denoted by  $\varepsilon_{\text{tag}}$, represents the fraction of particles that the tagging algorithm can identify, regardless of whether the identification is correct or not. 
The mistagging rate, denoted by $\omega_{\text{tag}}$, represents the proportion of incorrectly identified particles among those that are identified. The $\omega_{\text{tag}}$ is expressed as:
\begin{displaymath}
\omega_{\text{tag}} =
\frac{N_{\text{W}}}{N_{\text{R}} + N_{\text{W}}},
\end{displaymath}
where $N_{\text{R}}$ is the number of events correctly tagged, and $N_{\text{W}}$ is the number of events incorrectly tagged.
The difficulty in accurately identifying the initial flavor, along with the rate of misidentification, reduces the precision of extracting parameters from the fit.
Consequently, the effective sample size is reduced by a factor known as the tagging power, represented by $p$, in comparison to an ideal scenario of perfect tagging, where
\begin{displaymath}
	\label{eq:scaling}
	p = \varepsilon_\mathrm{tag}(1-2\omega_\mathrm{tag})^2.
\end{displaymath}

Another important factor that affects the precision of $\phis$ measurements is the resolution of the proper $\Bs$ decay time $t$  measurement, donated as $\sigma_t$. 
This resolution impacts the precision of $\phis$ measurements in the format of $\sigma({\phis}) \propto 1/\exp({-\frac12\Delta m_{s}^2\sigma_t^2})$, where $\Delta m_s$ is the mass difference of the two $B_s$ eigenstates, as detailed in Appendix.

A scaling factor, which is proportional to the $\sigma(\phis)$, can be established as follows:
\begin{equation}
\xi = \frac{1}{\sqrt{N_{b\bar{b}} \times \varepsilon} \times \sqrt{p} \times \exp\left(-\frac{1}{2}\Delta m_{s}^2\sigma_t^2\right)}.
\end{equation}

This scaling factor $\xi$ allows us to estimate the expected precision of $\phis$ in future $Z$-factories with
\begin{equation}
\sigma(\phis, \text{FE}) = \xi_{\text{FE}} \times \frac{\sigma(\phis, \text{EE})}{\xi_{\text{EE}}},
\end{equation}
where FE denotes a future experiment and EE denotes an existing experiment.

In this study, the precision and scaling factor for the existing experiment are estimated from the LHCb studies ~\cite{LHCb:2019nin}. For the LHCb measurement, the number of extracted signals is $N_{b\bar{b}}\times\varepsilon$ = \lhcbNBs. The flavor tagging power $p$ is \lhcbtag. The decay time resolution $\sigma_t$ is measured as \lhcbtime. The precision of $\phis$ is measured to be \lhcbRes. Consequently, the scale factor $\xi_{\text{lhcb}}$ is calculated to be \lhcbxi, and the ratio $\sigma(\phis)/\xi_{\text{lhcb}}$ is $\lhcbratio$.

The scaling factor for the future $Z$-factory is estimated through a Monte Carlo study. The details of the estimation will be elaborated in the subsequent sections.

% The scaling factor of the experiments at the High-Luminosity LHC is also estimated for comparison. Assuming no significant changes in detector acceptance and efficiency, tagging power, and decay time resolution at HL-LHC, the scaling factor is calculated by scaling the luminosity. At HL-LHC, the expected luminosity is $300~\mathrm{fb^{-1}}$, with respect to $1.9~\mathrm{fb^{-1}}$ at the current measurement of LHCb. The scaling factor is then $\xi_{\text{ HL-LHC -LHCb}}=\hllhcbxi$ and the expected precision is $\sigma(\phi_s,\text{ HL-LHC -LHCb})=\xi_{\text{ HL-LHC -LHCb}}\times\sigma(\phis)/\xi_{\text{lhcb}} = 3.2~ \mathrm{mrad}$.
The scaling factor for experiments conducted at the High-Luminosity LHC (HL-LHC) is also calculated for comparison. It is assumed that there will be no significant changes in the detector's acceptance, efficiency, tagging power, or decay time resolution at the HL-LHC. The scaling factor is determined by scaling for the increase in luminosity. At the HL-LHC, the anticipated luminosity is $300~\mathrm{fb^{-1}}$, compared to the current measurement of $1.9~\mathrm{fb^{-1}}$ at LHCb. 
% Additionally, it is expected that the cross-section will increase by a factor of $14/13$ when the collision energy is raised from $13~\mathrm{TeV}$ to $14~\mathrm{TeV}$.
Therefore, the scaling factor is \\ $\xi_{\text{HL-LHC-LHCb}}=0.0015$. Based on this, the expected precision for $\sigma(\phi_s,\text{HL-LHC-LHCb})$ is calculated to be \\ $\xi_{\text{HL-LHC-LHCb}} \times \frac{\sigma(\phis)}{\xi_{\text{lhcb}}} = 3.3~\mathrm{mrad}$. This estimate suggests a slightly more promising outcome than the one presented in Ref.~
\cite{LHCb:2018roe}, which is $4~\mathrm{mrad}$. 
The estimation presented in Ref.~\cite{LHCb:2018roe} is based on a projection from Ref~\cite{LHCb:2017hbp}.
The discrepancy between the two estimates could be attributed to the improvement of flavor tagging employed in the study of Ref.~\cite{LHCb:2019nin}, which marks an advancement over the methodologies used in the earlier study, Ref.~\cite{LHCb:2017hbp}.

The expected precision for the parameters $\Delta \Gamma$ and $\Gamma_s$ is estimated in a similar manner as the estimation of $\phis$ measurement precision.
These parameters are primarily influenced by the shape of the exponential decay and are less impacted by oscillatory behavior. Consequently, when the decay time resolution is small, they are not as sensitive to the tagging power and the resolution of the proper decay time, which distinguishes them from $\phis$ measurements. This assertion is further confirmed by simulations, as detailed in Appendix.
The variable
\begin{equation}
	\label{eq:scaling2}
	\zeta = 1/\left(
	\sqrt{N_{b\bar{b}}\times\varepsilon}
	\right)
\end{equation} is introduced as the scaling factor for $\Gamma_s$ and $\Delta\Gamma_s$. The scaling factor for LHCb is $\zeta_{\text{lhcb}} = 2.9\times10^{-3}$, estimated from Ref.~\cite{LHCb:2019nin}.

\subsection{CEPC and the baseline detector}
The CEPC and the baseline detector (CEPC-v4)~\cite{cepc2018cepc} are taken as an example to study the precision of $\phis$, $\Delta\Gamma_s$ and $\Gamma_s$.
As a baseline, the CEPC is assumed to run in the Tera-$Z$ mode, i.e., produces $10^{12}$ $Z$ bosons during its lifetime. 
The CEPC baseline detector consists of a vertex system, a silicon inner tracker, a TPC, a silicon external tracker, an electromagnetic calorimeter, a hadron calorimeter, a solenoid of 3 Tesla, and a Return Yoke.

\subsection{Monte Carlo sample and reconstruction}
\label{sec:MC}
A Monte Carlo signal sample is generated to analyze the geometric acceptance and the reconstruction efficiency of the $\BsJPsiPhi$ decay. Additionally, this sample is also used for the examination of the proper decay time resolution for the $B_{s}$, which has a direct correlation with the spatial resolution at the $B_{s}$ decay vertex.

Using the WHIZARD~\cite{whizard} generator, roughly 6000 events of $Z\rightarrow b\bar{b} \rightarrow B_{s}(\bar{B}_{s})+X$ are simulated. The $B_s(\bar{B}_{s})$ particles are then forced to decay through the $B_s(\bar{B}_{s})\rightarrow J/\psi(\rightarrow \mu^+ \mu^-)\phi(\rightarrow K^+ K^-)$ process using PYTHIA~8~\cite{pythia}, with a uniform distribution in phase space.

The simulation of particle transport within the detector utilizes MokkaC, the simulation software for the CEPC study, based on the GEANT~4~\cite{agostinelli2003geant4}. Based on Monte Carlo truth data, the reconstructed particles are categorized into hadrons, muons, and electrons.

The $J/\psi$ candidates are reconstructed from every pairing of a positively charged muon with a negatively charged muon. They are then selected based on the invariant mass window, ranging from $\jpsiml$ to $\jpsimr$~$\mathrm{GeV}/c^2$.
The $\phi$ candidates are reconstructed from every possible combination of a positively charged kaon and a negatively charged kaon. 
The $\phi$ candidates are selected within the mass window from \phiml~ to \phimr~$\mathrm{GeV}/c^2$. 
The $B_s(\bar{B}_{s})$ meson is reconstructed by combining all pairs of $J/\psi$ and $\phi$ candidates identified in the preceding steps. The four-momentum of the $B_s(\bar{B}_{s})$ meson is determined using the four-momentum of the $J/\psi$ and $\phi$, ensuring conservation of energy and momentum.
They are selected within a mass window ranging from $\bsml$ to $\bsmr$~$\mathrm{GeV}/c^2$. Following the reconstruction of the $B_s(\bar{B}_{s})$ meson, a decay vertex is constructed using the tracks associated with the $B_s(\bar{B}_{s})$.

As the CEPC was initially designed as a Higgs factory, the secondary vertex reconstruction algorithm and the flavor tagging algorithm are not in the standard CEPC software chain. A vertex reconstruction procedure and a simple flavor tagging algorithm were specially implemented for this study, which will be described in Sect.~\ref{sec:decaytimeresolution}.
% The signal sample is also used to estimate the flavor tagging power. For this purpose, only truth-level information is passed to the flavor tagging algorithm, i.e., assuming that we have perfect particle identification.

An additional sample of $Z\rightarrow b\bar{b}\rightarrow X$ is generated to show that a low background level is achievable through appropriate event selection criteria for the $\phis$ measurement. 
The detector simulation and event reconstruction processes are consistent with those applied to the signal sample.

\subsection{Statistics and acceptance $\times$ efficiency}
Assuming that all $b\bar{b}$ events can be selected with high purity, the background events in this work are the $b\bar{b}$ events that do not contain $\Bs\rightarrow J/\psi(\rightarrow\mu^+\mu^-)\phi(\rightarrow K^+ K^-)$ signal. 
The branching fraction of $b\bar{b}$ hadronized to $B_s$ is 10\%~\cite{ParticleDataGroup:2020ssz}.
The branching ratio of $\BsJPsiPhi$ is $1.08\times10^{-3}$. And the branching ratio of $J/\psi\rightarrow\mu^+\mu^-$, $\phi\rightarrow K^+ K^-$ are $6\%$ and $50\%$ respectively~\cite{ParticleDataGroup:2020ssz}. If the background is not suppressed by any event selection criteria, the number of background events is $1/(10\%\times 1.08\times 10^{-3} \times 6\% \times 50\%) = 3.1\times10^{5}$ times larger than the number of signal events. 
Applying the invariant mass selection criteria described in Sect.~\ref{sec:MC} to the background sample, the probability of reconstructing a fake $B_s$ candidate from a $Z\rightarrow b\bar{b} \rightarrow X$ event is $6.7\times10^{-6}$. Therefore, after the event selection, background statistics are of the same magnitude as the signal statistics.

\begin{figure}[!hbt]
\begin{center}
\includegraphics[width=0.49\textwidth]{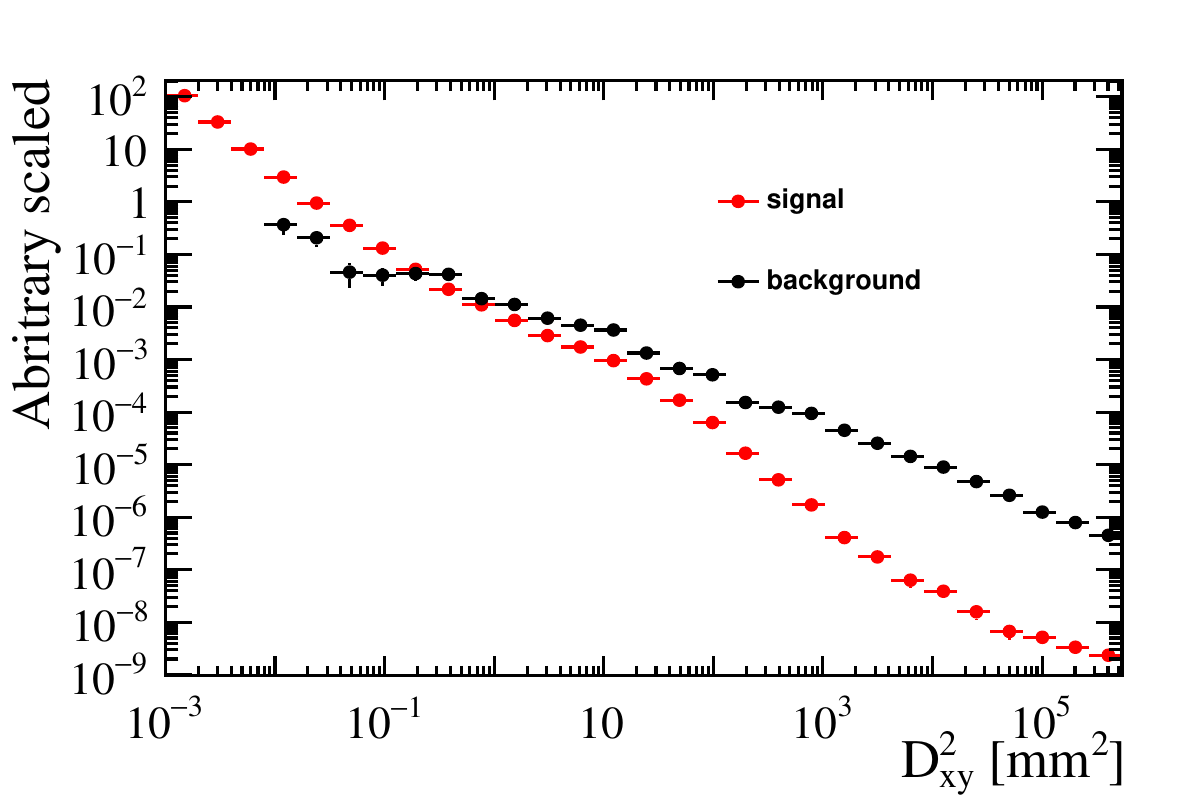}
\end{center}
\caption{$D_{xy}^2$ distributions of the signal and background.}
\label{fig:chi2}
\end{figure}

The combinatorial background events that pass the invariant mass selection criteria are further suppressed by using vertex information. In the background events, the fake $B_{s}$ candidates come from four arbitrarily combined tracks, two of which are lepton tracks and two of which are hadron tracks. The lepton usually has a large impact parameter, and the hadron has a small one. It is difficult to reconstruct a high-quality vertex with arbitrarily combined tracks. 
The $D_{xy}^2$ is used to measure the quality of the vertex reconstruction, where
\begin{displaymath}
D_{xy}^2 = \sum_\mathrm{tracks} d_{xy}^2.
\end{displaymath}
The $d_{xy}$ in the formula represents the distance from the reconstructed vertex to the track in the plane perpendicular to the beam direction. 
The $D_{xy}^2$ distributions of the signal and background are shown in Fig.~\ref{fig:chi2}.
% In both of the two plots, the rightmost bin with maximum $D_{xy}^2$ contains all the events with $D_{xy}^2 > 0.1~\mathrm{mm^2}$.
The vertex $D_{xy}^2$ of signal is usually very small. 
And the $D_{xy}^2$ of background is distributed over an extensive range. 
With a very loose cut at $D_{xy}^2 < 0.1~\mathrm{mm^2}$, $95\%$ of the signals are selected and $99.2\%$ of the backgrounds are discarded.

By employing a combination of invariant mass and vertex cut, the acceptance $\times$ efficiency of the signal is 75\%, while the background is maintained at 1\% of the signal level.

Due to potential particle misidentification, a small peaking background may be present in the signal region. Implementing a strict threshold on the hadron ID can reduce this peaking background; however, it would also result in diminished efficiency. This loss of efficiency is not considered in the present analysis due to the excellent PID performances of the CEPC.

At CEPC, the electron tracking performance is as good as that of muon tracking. The $J/\Psi$ could be reconstructed via the $J/\Psi\rightarrow e^+e^-$ channel as well. Consequently, the total effective sample size is considered to be roughly twice as much as when only the $J/\Psi\rightarrow \mu^+\mu^-$ decay channel is considered.

\subsection{Flavor tagging}
\subsubsection{Flavor tagging algorithm}
A simple algorithm is developed to identify the initial flavor of the particle. 
The idea of the algorithm is as follows: 

The $b$($\bar{b}$) quarks are predominantly produced in $b\bar{b}$ pairs that fly in the opposite directions because of the momentum conservation. The flavor of the opposite $b$-quark can be used to determine the initial flavor of the interested $B_s$. To judge the flavor of this opposite $b$-quark, we take a lepton and a charged kaon with maximum momentum in the opposite direction of the $B_s$. The lepton and kaon charge provides the flavor of the opposite $b$-quark.
Furthermore, when the $b$ quark is hadronized to a $B_s$ meson, another $s$ quark is spontaneously created, which has the chance to become a charged kaon, flying in a similar direction to the $B_s$. Based on this kaon, one can identify the flavor of the particle. The algorithm takes the leading particles (particles with the largest momentum projected onto the direction of the $B_s$). If these particles provide different determinants for the flavor, the algorithm says that it cannot identify the flavor.
The kaons and the muons from the $\BsJPsiPhi$ decay are excluded from the consideration in the above algorithms.

\subsubsection{Flavor tagging power}
The algorithm is applied to a Monte Carlo truth-level simulation, assuming perfect particle identification. 
The probability of finding a charged kaon at the near side (the angle between the momentum of the two particles is less than $\pi/2$) of the \Bs is $56.6\pm1.2\%$. Within the events with near-side kaons, $79.8\pm1.4\%$($20.2\pm1.4\%$) of the leading kaons are $K^+$($K^-$) % [FIXME: check the sign] 
if a $\Bs$ rather than $\bar{B}_s$ is produced. The significant difference between the abundance of $K^+$ and $K^-$ makes the nearside kaon a powerful distinguish observable to identify the initial flavor of the $B_s$ meson. At the opposite side (the angle between the momentum of the two particles is larger than $\pi/2$), the probability of finding a charged kaon is $72.5\pm1.1$\%. The percentage of the $K^+$ is $31.2\pm1.4\%$, while the percentage of the $K^-$ is $68.8\pm1.4\%$ for the leading kaons. The probability of finding an electron or muon at the opposite side of $\Bs$ is $38.3\pm2.0\%$, where the probability of the leading particle to be an electron, positron, muon, or anti-muon is $33.8\pm2.0\%$, $22.5\pm1.7\%$, $26.4\pm1.8\%$ and $17.3\pm1.6\%$, respectively.

Based on the particles detected in the events, each of the three tagging discriminators (opposite kaon, opposite lepton, and same-side) yields a determination regarding the flavor of the produced $\Bs$ mesons.
These determinations are then classified as either $B_s$, $\bar{B}_s$, or undetermined, which correspond to a voting score of $1$, $-1$, or $0$, respectively. Outcomes identified as $B_s$ or $\bar{B}_s$ are classified as definitive decisions.
The voting scores from the three tagging discriminators are added. The initial flavor of the meson is then inferred from the sum's sign: a positive sum signifies $B_s$, a negative sum signifies $\bar{B}_s$, and a zero sum denotes an indeterminate flavor.
In $5.9\pm0.4\%$ of instances, all three discriminators render a definitive decision, while in $22.9\pm0.7\%$ of cases, none of the discriminators are able to provide a definitive decision.
In $43.7\%\pm0.9\%$ of instances, a single discriminator gets a definitive decision. Conversely, in $27.4\%\pm0.8\%$ of cases, two discriminators concur on a definitive tagging decision. Within this subset, $62.4\%\pm1.7\%$ of the time, both discriminators agree on the same decision.
The final tagging efficiency is estimated as $66.8\pm0.9\%$. The mistagging rate is $22.5\pm0.9\%$. The tagging power is estimated to be $20.2\pm1.4\%$.
% , the fraction of events the tagging algorithm can provide a tag decision, 

\begin{figure}[!hbt]
\begin{center}
\includegraphics[width=0.49\textwidth]{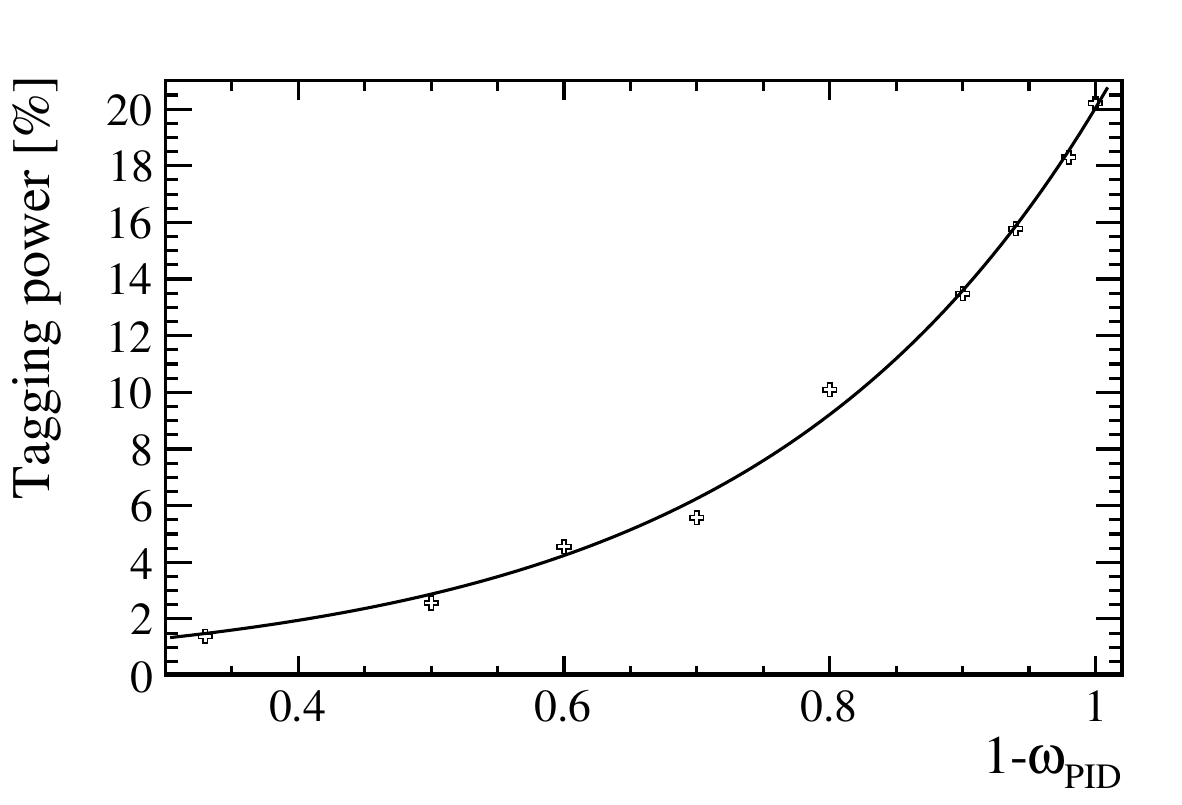}\end{center}
\caption{Tagging power as a function of the correct particle identification rate~$1-\omega_\mathrm{PID}$.}
\label{fig:pid_tag}
\end{figure}

Additionally, if the particle identification is imperfect, the flavor tagging power decreases. This impact is analyzed by deliberately mislabeling hadrons with incorrect IDs. A pion is mislabelled as either a kaon or a proton with a probability of $\omega_{\mathrm{PID}}/2$ for each. This method of random mislabeling is similarly applied to kaons and protons.
The tagging power varying with the correct particle identification rate $1-\omega_\mathrm{PID}$ is shown in Fig.~\ref{fig:pid_tag}. The tagging power is sensitive to the $\omega_\mathrm{PID}$ parameter. At the region of $1-\omega_\mathrm{PID}\sim 0.33$, where the particle identification ability is totally missing, the tagging power is degraded to around 0.

It is also worthwhile to explore how the tagging power is degraded in a realistic scenario.
The momentum-dependent particle identification on CEPC was investigated in a previous study~\cite{Zhu:2022hyy}.
The momentum-dependent $p/K/\pi$ separation power is applied in this study to simulate the hadron misidentification. 
The seperation power $\langle S\rangle$ quoted from~\cite{Zhu:2022hyy} is used in the following way:
For instance, to assign an ID to a $\pi$, We generate a random variable by employing a Gaussian distribution with a mean of 0 and a standard deviation of 1. If the generated random number is less than $\langle S\rangle_{K/\pi}/\sqrt{2}$, the ID is assigned to $\pi$ as $\pi$. Conversely, if it is greater than $\langle S\rangle_{K/\pi}/\sqrt{2}$, it is assigned as $K$. 
Likewise, the same procedures are applied to $p$.
Additionally, the $K$ could be mistaken for either $p$ or $\pi$. If the generated random number is less than $-\langle S\rangle_{K/\pi}/\sqrt{2}$, 
the ID of the particle is assigned as $\pi$. 
While if the random number is larger than $\langle S\rangle_{K/p}/\sqrt{2}$, the ID of $p$ is assigned to this kaon.
The particles with assigned IDs are utilized to tag the initial $B_s$ flavor according to the tagging algorithm that was previously described.
Under the intrinsic case, without considering the effects of the readout electronics, the tagging power is 19.1\%. In a more realistic case, if the particle identification resolution is degraded by 30\%, corresponding to a reduction of $\langle S\rangle$ by 30\%~\cite{An:2018jtk}, the tagging power becomes 17.4\%. The decrease of the tagging power with a worse PID performance is because the large difference between the abundance of $K^+$ and $K^-$ is smeared by the misidentified $\pi^+$ and $\pi^-$.

\

\subsection{Decay time resolution}
\label{sec:decaytimeresolution}
The precision of $\phis$ is affected by the inaccurate determination of the decay time. The proper decay time for the $B_s$ meson is determined using the decay vertex position and the transverse momentum of the $B_s$ as follows:
\begin{equation}
	t = \frac{m_s r}{p_{\mathrm{T}}},
 \label{eq:decaytime}
\end{equation}
where $r = \sqrt{x^2+y^2}$ represents the $B_s$ decay vertex position in the transverse plane, $p_{\mathrm{T}}$ denotes the transverse momentum of the $B_s$ meson, and $m_s$ represents the mass of the $B_s$. The $B_s$ decay vertex is constructed from the four tracks from the $B_s$ decay. It is assumed that the primary vertex, the production point of the $B_s$, is located at the origin. The resolution for determination of the primary vertex is considered negligible, given that the abundance of tracks available to reconstruct the primary vertex far exceeds those available for determining the $B_s$ decay location.

The decay point of the $B_s$ meson is determined by minimizing the $\chi^2$ value, which is calculated as the sum of the squares of the shortest distances from the vertex to each of the four helical tracks. This technique is an adaptation of the method described in Ref.~\cite{Fruhwirth:2020zbo}.
 
To simplify the minimization process, the helical track is approximated as a straight line that is tangent to the helix at point $ P^{(0)}$. This point $P^{(0)}$ is the closest to a reference point on the helix. 
In this algorithm, the true decay vertex of $B_s$ is used as the reference point, denoted by $v_{\text{ref}}$.

Subsequently, each of the tracks is parameterized by a point $r_i$ and a direction $a_i$.
Consequently, the minimization process is replaced by the solving of a matrix equation
\begin{equation}
    v = v_\mathrm{ref} + (\sum H_i)^{-1}\cdot\sum r_i,
\end{equation}
where $v$ is the $B_s$ decay point position and  
\begin{displaymath}
    H = \begin{pmatrix}
    a_y^2 + a_z^2 && - a_x a_y && -a_x a_z \\
    -a_x a_y && a_x^2+a_y^2 && -a_y a_z \\
    -a_x a_z && -a_y a_z && a_x^2+a_y^2 \\
    \end{pmatrix}.
\end{displaymath}

\begin{figure}[!hbt]
\begin{center}
\includegraphics[width=0.49\textwidth]{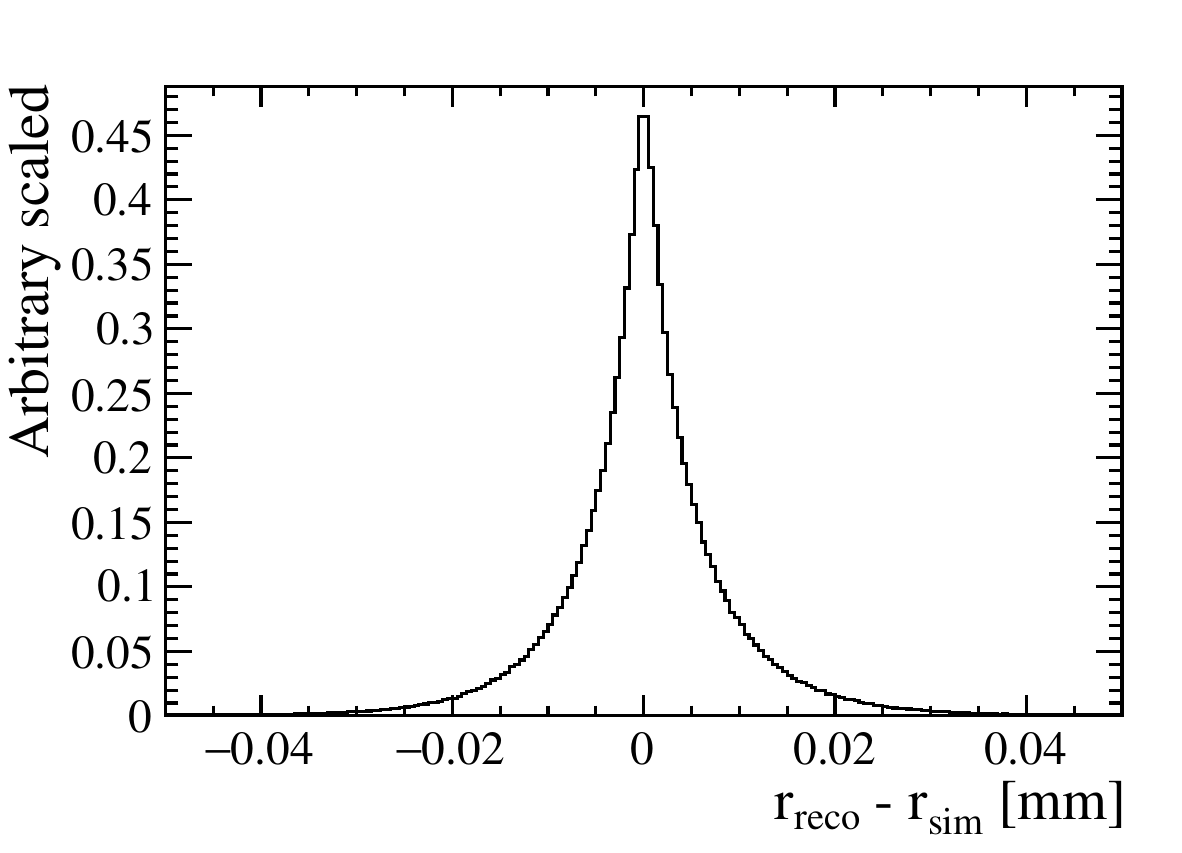}
\end{center}
\caption{Distribution of $r_\text{reco}-r_\text{sim}$.}
\label{fig:vtx_res}
\end{figure}

The difference between the reconstructed $B_s$ decay vertex position ($r_\mathrm{reco}$) and the truth vertex position ($r_\mathrm{sim}$) is shown in Fig.~\ref{fig:vtx_res}. 

\begin{figure}[!hbt]
\begin{center}
\includegraphics[width=0.49\textwidth]{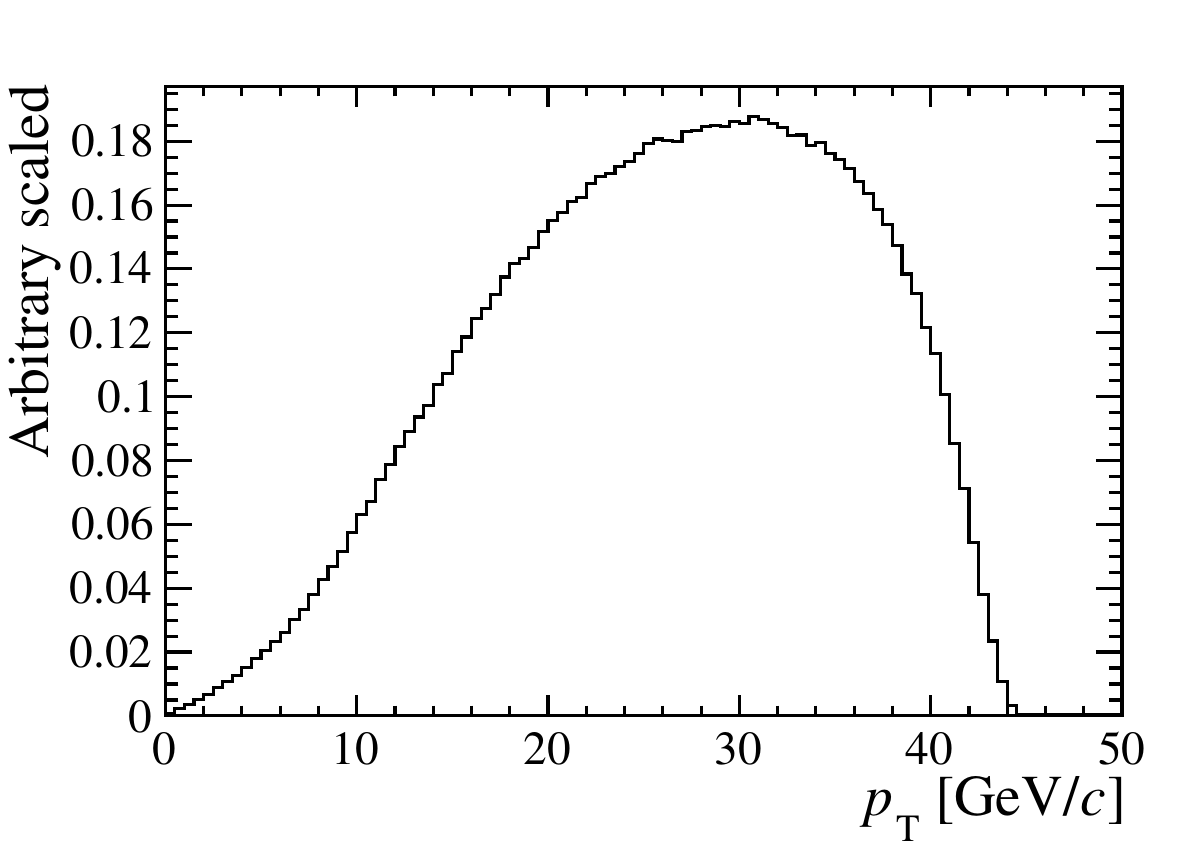}
\end{center}
\caption{Transverse momentum distribution of $B_s$.}
\label{fig:pt}
\end{figure}

The transverse momentum distribution of $B_s$ mesons is shown in Fig.~\ref{fig:pt}. The majority of these mesons have large transverse momentum because they hadronize from a high-energy $b$-quark, which carries nearly half of the beam energy. Consequently, a substantial transverse momentum corresponds to a large Lorentz boost factor, which substantially enhances the resolution of the proper decay time.

\begin{figure}[!hbt]
\begin{center}
\includegraphics[width=0.49\textwidth]{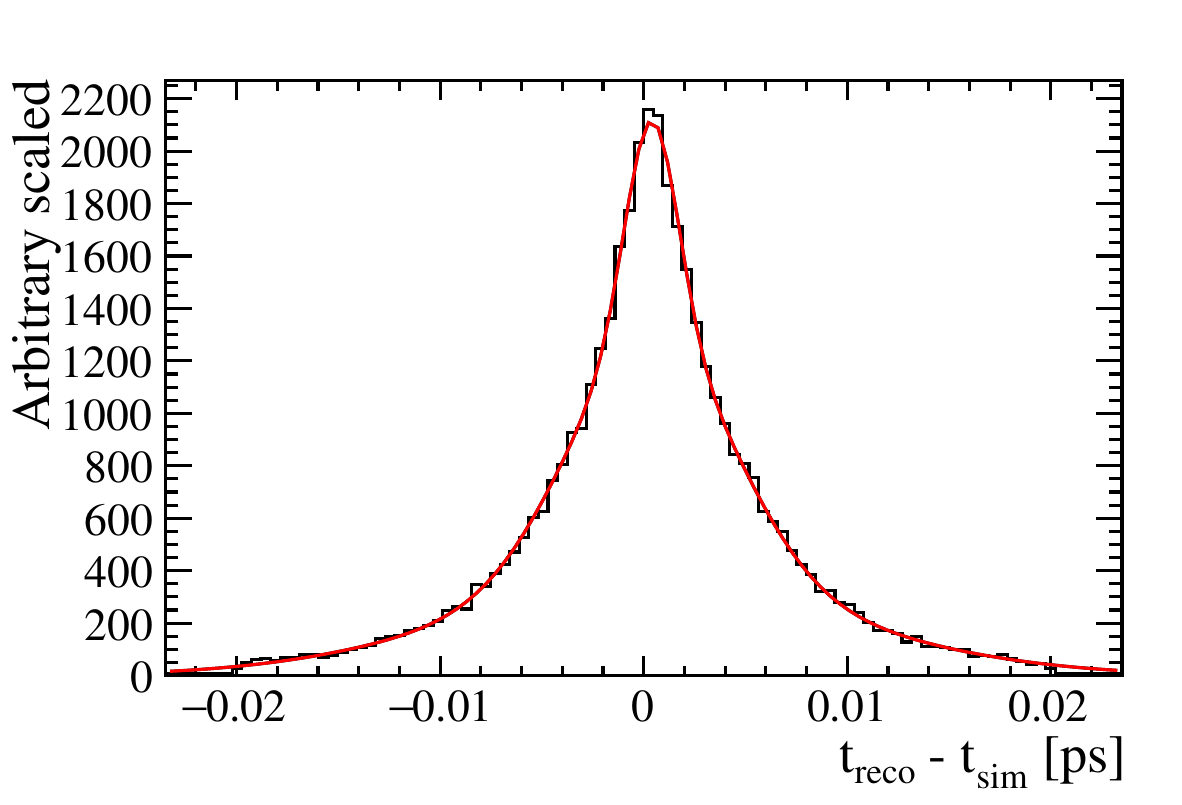}
\end{center}
\caption{Distribution of $t_\text{reco}-t_\text{sim}$. The distribution is fitted with the sum of three Gaussian functions with equal mean.}
\label{fig:time_res}
\end{figure}

Figure~\ref{fig:time_res} shows the distribution of the differences between $t_{\text{reco}}$ and $t_{\text{sim}}$ for the same event. Both $t_{\text{reco}}$ and $t_{\text{sim}}$ represent the proper decay time as calculated from Eq.~(\ref{eq:decaytime}). To derive $t_{\text{sim}}$, detector effects are not considered, and the vertex and $p_\mathrm{T}$ are obtained directly from the Monte Carlo truth record. In contrast, $t_{\text{reco}}$ is determined after the particles undergo detector simulation, as well as track and vertex reconstruction, using the reconstructed vertex position and transverse momentum to calculate the proper decay time.
The distribution is fitted using the sum of three Gaussian functions with the same mean value. 
The effective time resolution is combined as
\begin{displaymath}
\sigma_{t}=\sqrt{
-\frac{2}{\Delta m_s^2}\ln(\sum_i f_i e^{-\frac{1}{2}\sigma_i^2\Delta m_s^2})
},
\end{displaymath}
where $f_i$ and $\sigma_i$ are the fraction and width of the i-th Gaussian function. 
The effective resolution of the decay time, achieved through the combined effect of the three Gaussian resolution models, is ${4.7~\mathrm{fs}}$. 

% The CEPC detector is expected to achieve a proper time resolution around 10 times better than the current LHCb. For one reason, the CEPC has a excellent vertex detector and a precise tracker. For another reason, the $B_s$ produced from $Z$-decay is energetic resulting in a smaller proper time resolution. 

%The conclusion is robust and well-known.
% The difference of $t_{xy}$ calculated $t_{xy}$ calculated 
% Figure shows the vertex resolution and the decay time resolution. 

\begin{comment}
The \phis parameter appears in the $a_k, b_k, c_k, d_k$ parameters of the fitting function. 
 
And that terms are nearly linear with \phis(If they have \phis in the first order expansion.) So to study the effects on the parameter from the resolution of t. We did a simple MC study by fitting a function of $h = \phi_s \cos(\Delta m_s t)$. The resolution of $\phi_s$ follows $1/\exp(-\frac12 \Delta m_s^2 \sigma_t^2)$: 
\end{comment}

\begin{comment}
$\Delta m_s=17.757\times10^12  \hbar s^{-1}$

	LHCb: $\sigma_t=50 fs \rightarrow 0.67$

	CEPC: $\sigma_t=10 fs \rightarrow 1$

	It is a conservative assumption that the SV reconstruction is 10 times worse than PV reconstruction.
We also need that study from direct simulation.
\end{comment}

\subsection{Decay time acceptance}
The possible impact of non-uniform decay time-dependent efficiency, known as decay time acceptance, on the precision of $\Gamma_s(\Delta\Gamma_s)$ measurement has been evaluated. While it was considered that different time acceptance profiles at hadron and lepton colliders might markedly influence the precision of $\Gamma_s$ and $\Delta\Gamma_s$ measurements, our findings indicate that the effect is not substantial.

\begin{figure}[!hbt]
\begin{center}
\includegraphics[width=0.49\textwidth]{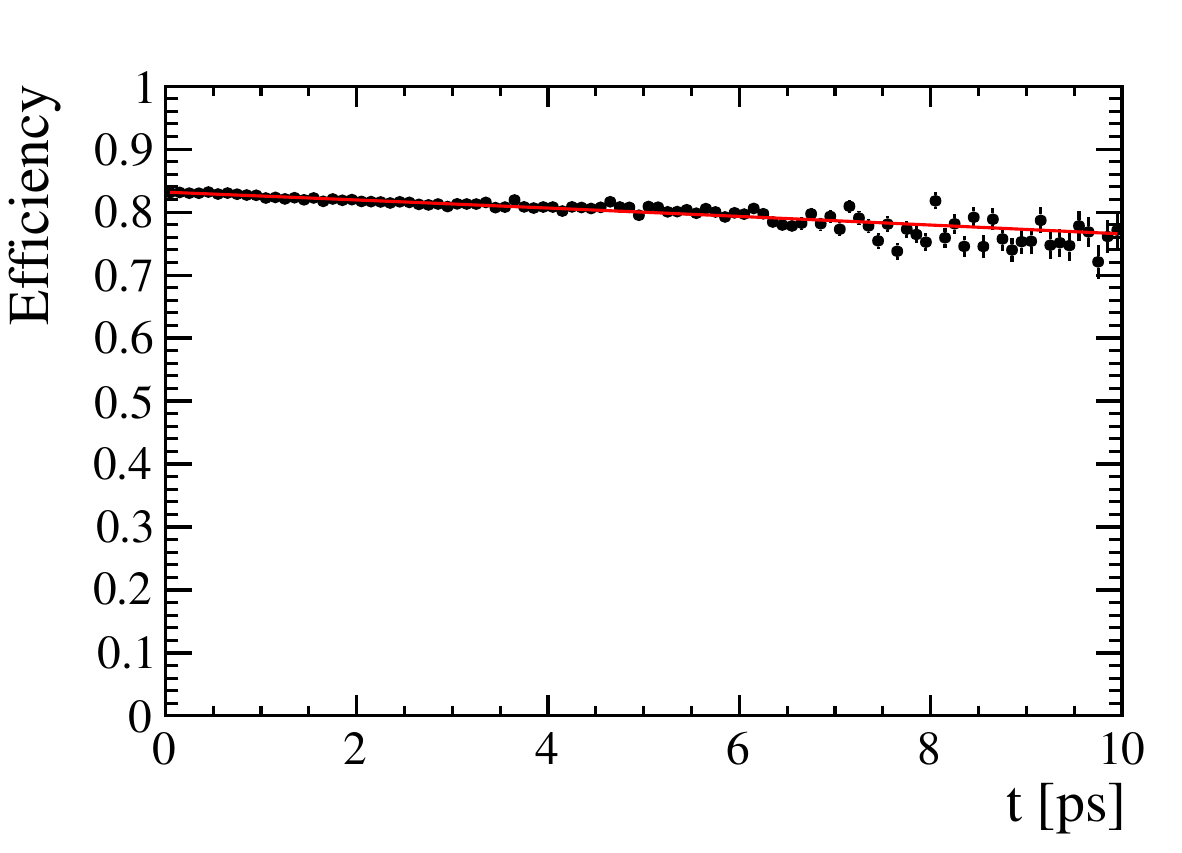}
\end{center}
\caption{Reconstruction efficiency as a function of the proper decay time.}
\label{fig:time_eff}
\end{figure}

Figure \ref{fig:time_eff} shows the reconstruction efficiency of the $B_s$ meson as a function of its proper decay time at the CEPC. The efficiency decreases at larger decay times due to the tracks of the $B_s$ with larger flight distances deviating increasingly from the interaction point, which complicates the reconstruction process.
The efficiency is parameterized with a 2nd-order polynomial function $f_\mathrm{acc}^\mathrm{CEPC} = 0.83 - 0.0061 \times t - 5.25 \times 10^{-5} \times t^2$. The time acceptance profile from LHCb, as referenced in Ref.~\cite{Liu:2786808} is $f_\mathrm{acc}^\mathrm{LHCb} = f_1 \times f_2$, where 
\begin{displaymath}
    f_1 = 1 -0.037 \times t + 0.001 \times t^2,
\end{displaymath}
and
\begin{displaymath}
    f_2 = \frac{[1.589\times(t-0.097)]^{1.150}}{1+[1.589\times(t-0.097)]^{1.150}}.
\end{displaymath}

A toy Monte Carlo simulation is employed to explore the impact of the time acceptance, as elaborated in the Appendix. It is found that, with the two distinct time acceptance profiles, the fitted parameters differ by only $17.7\%$, which will be neglected in the final results.

% 0.831674 -0.00610988 * t -5.25383e-05 * t^2
% CEPC 1.02581e-02
% 8.91698e-03
% 1.08631e-02

\section{Results}
\subsection{Precision of $\phis$, $\Gamma_s$ and $\Delta\Gamma_s$}

The above simulations show that in future $Z$-factories, the proper decay time resolution can reach $4.7~\mathrm{fs}$, the detector \\ acceptance$\times$efficiency can be as good as $75\%$, and the flavor tagging power can be $17.4\%$ under a conservative assumption on the PID performance.
In addition, the acceptance and efficiency is almost flat in decay time.
Assuming the future $Z$-factory operating in Tera-$Z$ mode (i.e., $10^{12}$ $Z$), the scaling factor $\xi_{FE}$ is $0.0021$. The expected $\phis$ resolution is $\sigma(\phis,\text{FE})=\xi_{\text{FE}}\times\sigma(\phi_s,\text{LHCb})/\xi_\text{LHCb} = 4.6~\mathrm{mrad}$.
% which is competitive to $3.3~\mathrm{mrad}$, the expected $\phis$ measurement resolution of LHCb at the HL-LHC. 

The $\Gamma_s$ and $\Delta\Gamma_s$ depend weakly on tagging power and decay time resolution. The 3.7 times better flavor tagging power and 1.92 times better time resolution factor of CEPC, in contrast to $\phis$, have negligible effects on these observables. The estimated resolution is $2.4~\mathrm{ns^{-1}}$ for $\Delta\Gamma_s$ and $0.72~\mathrm{ns^{-1}}$ for $\Gamma_s$. The measured resolution of $\Gamma_s-\Gamma_d = 0.0024~\mathrm{ps^{-1}}$\cite{LHCb:2019nin} is taken as the resolution of $\Gamma_s$.

\begin{figure}[!hbt]
    \begin{center}
\includegraphics[width=\textwidth]{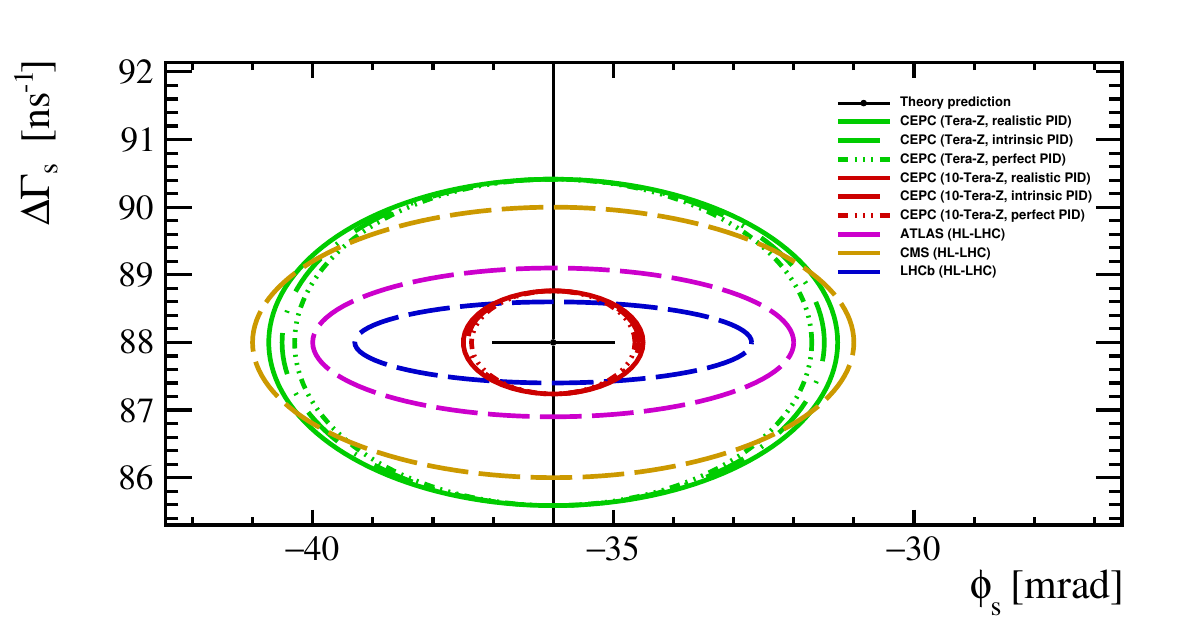}
    \end{center}
    \caption{Expected confidential region ($68\%$ confidential level) of $\Delta\Gamma_s-\phis$. The black point is the Standard Model prediction from CKMFitter group~\cite{Charles:2015gya} and HQE theory calculation~\cite{HQE}. The uncertainty of $\Delta\Gamma_s$ is $6~\mathrm{ns}^{-1}$. The green and red curves represent the expected precision of Tera-$Z$ CEPC and 10-Tera-$Z$ CEPC, respectively. 
    The blue dashed curve represents the LHCb at the HL-LHC, projected in this study. The magenta and yellow curves are the projections from ATLAS and CMS at the HL-LHC~\cite{ATLAS:2018gue,CMS:2018wpq}, respectively.
    All the circles are centered at the standard model central value.}
    \label{fig:phisgamma}
    \end{figure}

Figure~\ref{fig:phisgamma} shows the expected confidential range ($68\%$ confidential level) of $\Delta\Gamma_s-\phis$. The black dot is the prediction of the standard model from CKMFitter group~\cite{Charles:2015gya} and HQE theory calculation~\cite{HQE}. 
The green curves and red curves represent the expected precision of Tera-$Z$ CEPC and 10-Tera-$Z$ CEPC, respectively. The different line styles represent different PID performance assumptions, where the solid line represents the conservative PID performance assumption, which is a degradation of $30$\% of the intrinsic PID performance. The blue dashed curve represents the LHCb at the HL-LHC, projected in this study. The magenta and yellow curves are the projections from ATLAS and CMS at the HL-LHC~\cite{ATLAS:2018gue,CMS:2018wpq}, respectively.

The \phis resolution at the 10-Tera-$Z$ CEPC can reach the current precision of SM prediction. All the future experiment measurements of $\Delta\Gamma_s$ can provide stringent constraints on the HEQ theory. The CEPC could do a better job on the measurement of the $\phis$ than the measurement of the $\Delta\Gamma_s$ because the flavor tagging and decay time resolution are excellent.

\subsection{Penguin pollution}
If the penguin diagram is considered in the $B_s$ decay, the relation between $\phis$ and $\beta_s$ should be corrected as
\begin{equation}
\phi_s = -2 \beta_s + \Delta\phi_s(a,\theta).
\end{equation}
The shift $\Delta\phis$ could be expressed as
\begin{equation}
\tan(\Delta\phi_s)
=
\frac{2 \epsilon a \cos\theta\sin\gamma + \epsilon^2a^2\sin(2\gamma)}
{1+2\epsilon a\cos\theta\cos\gamma+\epsilon^2a^2\cos(2\gamma)},
\end{equation}
where $a$ and $\theta$ are penguin parameters, $\epsilon=\lambda^2/(1-\lambda^2)$ is defined through a Wolfenstein parameter $\lambda$, and $\gamma$ is the angle $\gamma$ of the Unitarity Triangle.
% =0.0536
% gamma != 73.2
% ~\cite{ParticleDataGroup:2020ssz}
% =65.9^\circ

Control channels, such as $B\rightarrow J/\Psi \rho$ and $\Bs\rightarrow J/\Psi K^*$, were employed to determine the penguin parameters $a$ and $\theta$, proposed by Ref.~\cite{Frings:2015eva}. 
In this study, the LHCb measurements involving $\Bs\rightarrow J/\Psi K^*$ are utilized to estimate the expected precision of $\Delta\phis$\cite{LHCb:2014xpr,LHCb:2015esn}. This is based on the assumption that the findings from the $\Bs\rightarrow J/\Psi \phi$ measurements can be directly applied to the $\Bs\rightarrow J/\Psi K^*$ analysis, despite the topological differences between the two decay channels.

The observables in $\Bs\rightarrow J/\Psi K^*$ measurements are
\begin{equation}
A^\mathrm{CP} = -
\frac{2 a \sin{\theta}\sin{\gamma}}
{1-2 a \cos{\theta}\cos{\gamma}+a^2},
\label{eq:ACP}
\end{equation}
and
\begin{equation}
H = \frac{1 - 2 a \cos{\theta}\cos{\gamma} + a^2}
{1 + 2 \epsilon a \cos{\theta}\cos{\gamma} + \epsilon^2 a^2},
\label{eq:H}
\end{equation}
where $A^\mathrm{CP}$ is the $CP$ asymmetry and $H$ is an observable constructed containing the branching fraction information,
assuming the $\mathrm{SU}(3)$ symmetry.

\begin{figure}[!hbt]
    \begin{center}
\includegraphics[width=0.5\textwidth]{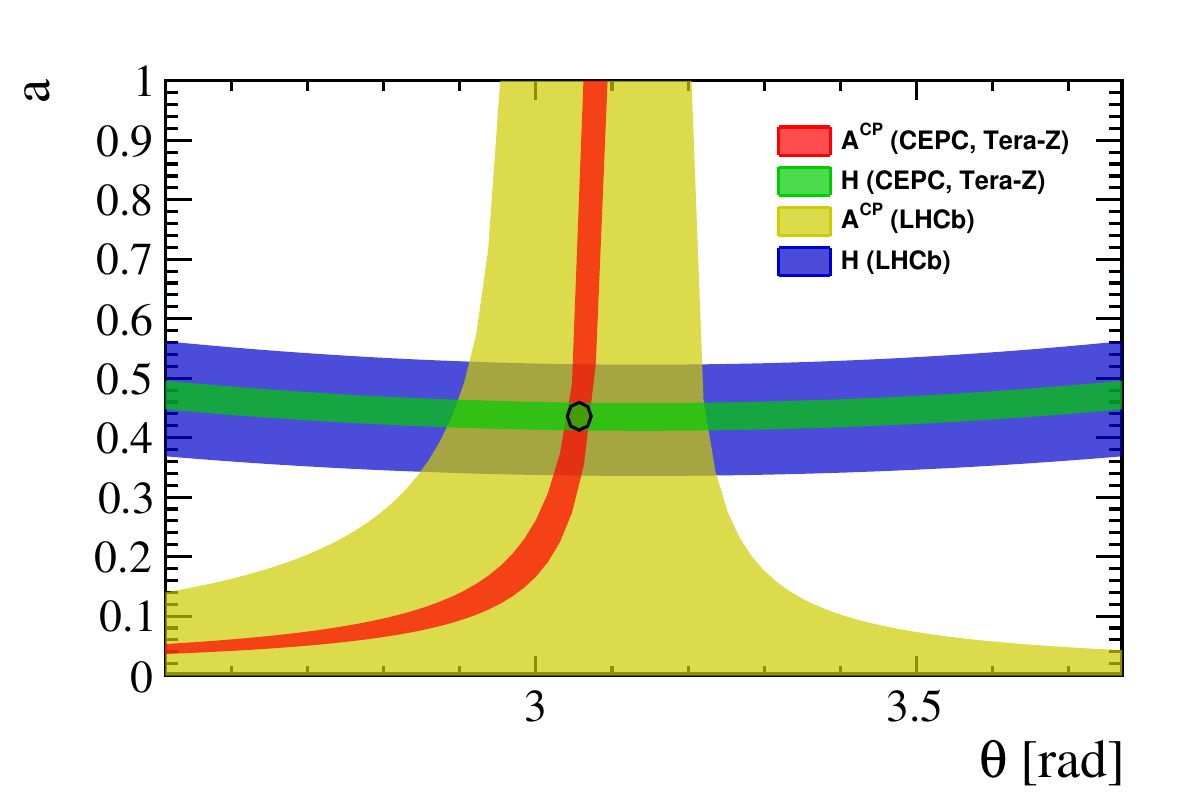}
    \end{center}
    \caption{The constraints on the penguin parameters $a$ and $\theta$ arise from $A^\mathrm{CP}$ and $H$. The black contour shows the anticipated one standard deviation ($1 \sigma$) limit resulting from the $A^\mathrm{CP}$ and $H$ constraints at CEPC.}
        \label{fig:penguin}
\end{figure}

The parameters $A^\mathrm{CP}$ and $H$ are polarization-dependent. The transverse components are measured at LHCb as
\begin{displaymath}
A^\mathrm{CP}_{\perp} = -0.049\pm0.096, H_{\perp} = 1.46\pm0.14.
\end{displaymath}

The constraints on the penguin parameters $a$ and $\theta$, as defined by Eq.~\ref{eq:ACP} and Eq.~\ref{eq:H}, and within the ranges of $A^\mathrm{CP}_{\perp} = -0.049 \pm 0.096$ and $H_{\perp} = 1.46 \pm 0.14$, are shown in Fig.~\ref{fig:penguin} as blue and yellow bands, respectively.
At future Tera-$Z$ $Z$-factory, the $H$ is expected to improve according to the Eq.~(\ref{eq:scaling2}), while the $A^\mathrm{CP}$ improves according to the Eq.~(\ref{eq:scaling}).  
Therefore, the $A^\mathrm{CP}$ and $H$ are expected to be measured at CEPC with the precision
\begin{displaymath}
    \sigma(A^\mathrm{CP}) = 0.0090, \sigma(H) = 0.035,
\end{displaymath}
taking into account the expected improvement of LHCb Run 2 compared with LHCb Run 1.
The anticipated constraints on the penguin parameters $a$ and $\theta$ at CEPC are represented by green and red bands in Fig.~\ref{fig:penguin}, with the central values of $A^\mathrm{CP}$ and $H$ from LHCb measurements.
The expected uncertainty of $a$ and $\theta$ is obtained by a $\chi^2$ fit to Eq.~(\ref{eq:ACP}) and~(\ref{eq:H}), resulting in
\begin{displaymath}
    a = 0.436\pm 0.023, \theta = 3.057\pm0.016^\circ.
\end{displaymath}
The black contour in Fig.~\ref{fig:penguin} outlines the region that falls within one standard deviation of the fitted value.

With an error propagation neglecting the correlation between $a$ and $\theta$, the precision of the penguin shift is estimated as 
$\sigma(\Delta\phi_s) = 2.4~\mathrm{mrad}$.

\begin{figure}[!hbt]
    \begin{center}
\includegraphics[width=0.5\textwidth]{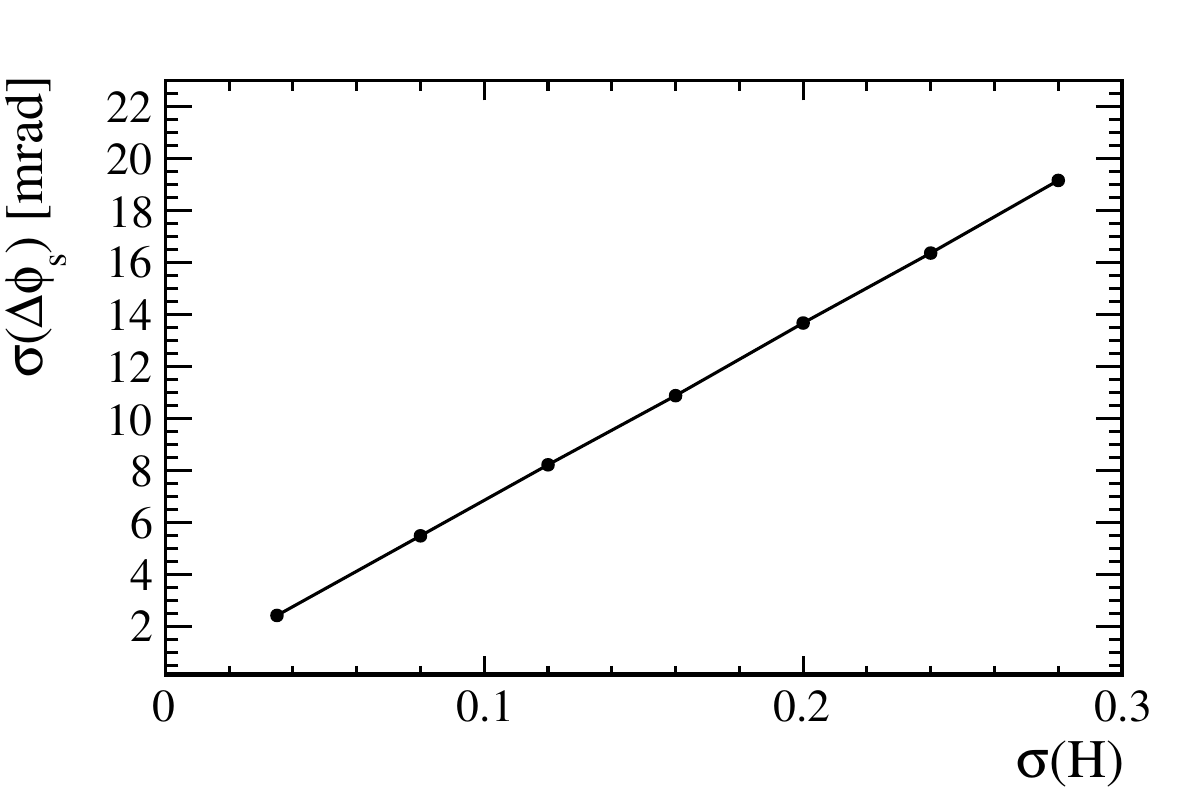}
    \end{center}
\caption{The variation of $\sigma(\Delta\phi_s)$ with respect to $\sigma(H)$.
To obtain $\sigma(\Delta\phi_s)$, the expected resolution of $\sigma(A^\mathrm{CP}) = 0.0090$ at CEPC is used.}
\label{fig:penguin2}
\end{figure}

However, the $\mathrm{SU}(3)$ symmetry does not always hold, and thus controlling $\sigma(H)$ requires additional theoretical efforts. The degradation of $\sigma(\Delta\phi_s)$ alongside the degradation of $\sigma(H)$ is shown in Fig.~\ref{fig:penguin2}.
To obtain $\sigma(\Delta\phi_s)$, the expected resolution of $\sigma(A^\mathrm{CP}) = 0.0090$ at CEPC is used. The procedure for determining $\sigma(\Delta\phi_s)$ with different $\sigma(H)$ follows the same methodology as the aforementioned statements.
The rightmost point on the Fig.~\ref{fig:penguin2} corresponds to $\sigma(H) = 0.28$, which reflects the theoretical uncertainty from the calculations in Ref.~\cite{DeBruyn:2048174}. 
It is demonstrated that $\sigma(\Delta\phi_s)$ is roughly linearly dependent on $\sigma(H)$, and clearly, without improved theoretical input, the control of penguin contamination will be far from satisfactory.

\section{Summary}
\begin{table*}[t]
\begin{center}
\caption{Parameters table of factors to calculate the precision of $\phi_s$, $\Gamma_s$ and $\Delta\Gamma_s$. The terms with $*$ means that the factor is insensitive to the resolution of $\Gamma_s$ and $\Delta\Gamma_s$.}
\label{tab:results}
%\begin{tabular}{cccccc}
\begin{tabular}{@{\extracolsep{\fill}}cccccc}
    \hline
    & LHCb (HL-LHC) & CEPC (Tera-Z) & CEPC/LHCb& \\
    \hline
    $b\bar{b}$ statics &$43.2\times10^{12}$&$0.152\times10^{12}$
    &1/284&& \\
    Acceptance$\times$efficiency &$7\%$&$75\%$&10.7&&\\
    Br &$6\times10^{-6}$&$12\times10^{-6}$&2&& \\
    Flavour tagging${}^*$ &$4.7\%$&$17.3\%$&{3.7}&& \\
    Time resolution${}^*$ ($\exp({-\frac12\Delta m_{s}^2\sigma_t^2})^2$) &$0.52$&$1$&1.92&& \\
    \hline
    $\sigma_t (\mathrm{fs})$ & 45& 4.7& \\
    \hline
    scaling factor $\xi$ &0.0015&0.0021&1.4&& \\
    $\sigma(\phis)$ &$3.3~\mathrm{mrad}$&$4.6~\mathrm{mrad}$&&& \\
    \hline
    \end{tabular}
    \end{center}    
\end{table*}

It is found that operating at Tera-$Z$ mode, the expected precision can reach: $\sigma(\phis) = 4.6~\mathrm{mrad}$, $\sigma(\Delta\Gamma_s) = 2.4~\mathrm{ns^{-1}}$ and $\sigma(\Gamma_s) = 0.72~\mathrm{ns^{-1}}$.
    As shown in Table~\ref{tab:results}, the statistical disadvantage of the Tera-$Z$ $Z$ factory can be compensated with a much cleaner environment, good particle identification, and accurate track and vertex measurement. 
    Without flavor tagging and time resolution benefits, the $\Gamma_s$ and $\Delta\Gamma_s$ resolution are much worse than expected for the LHC at high luminosity. Only with the 10-Tera-$Z$ $Z$ factory can the expected resolution of $\Delta\Gamma_s$ and $\Gamma_s$ be competitive. 
    % This represents a requirement for the future $Z$-factory. 
With the $B_s\rightarrow J/\Psi K^*$, considering only the transverse component, the penguin shift is expected to be measured as a precision of $\sigma(\Delta\phis) = 2.4~\mathrm{mrad}$. Controlling the penguin pollution is feasible, provided that the theoretical uncertainty is managed effectively.

The study presents clear performance requirements for detector design. The flavor tagging algorithm currently relies only on the leading particle information, suggesting there is potential to refine the algorithm further to improve the precision of $\phis$ measurements. A tagging power of $\sim 30\%$ is forseenable according to the experiences from the Ref.~\cite{Belle:2004uxp}. Particle identification is critical; the performance of tagging degenerates fastly when particle misidentification occurs. Distinguishing between different hadrons using particle identification data enables more precise event selection. Moreover, robust vertex reconstruction is essential to suppress combinatorial background. While the present decay time resolution is satisfactory, further improvements in time resolution are unlikely to increase the precision of $\phis$ measurements.

\appendix
\section*{Appendix}
\label{sec:appendix}
\begin{figure}[!hbt]
\begin{center}
\includegraphics[width=0.49\textwidth]{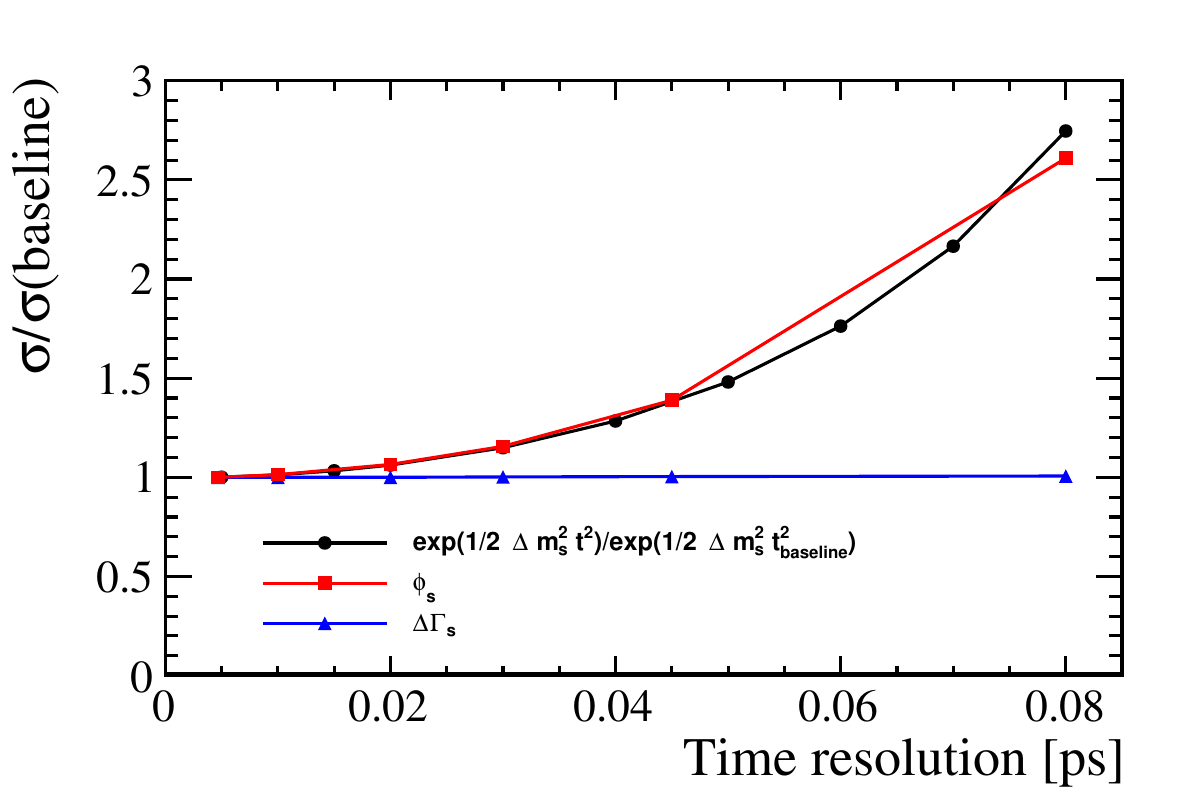}
\includegraphics[width=0.49\textwidth]{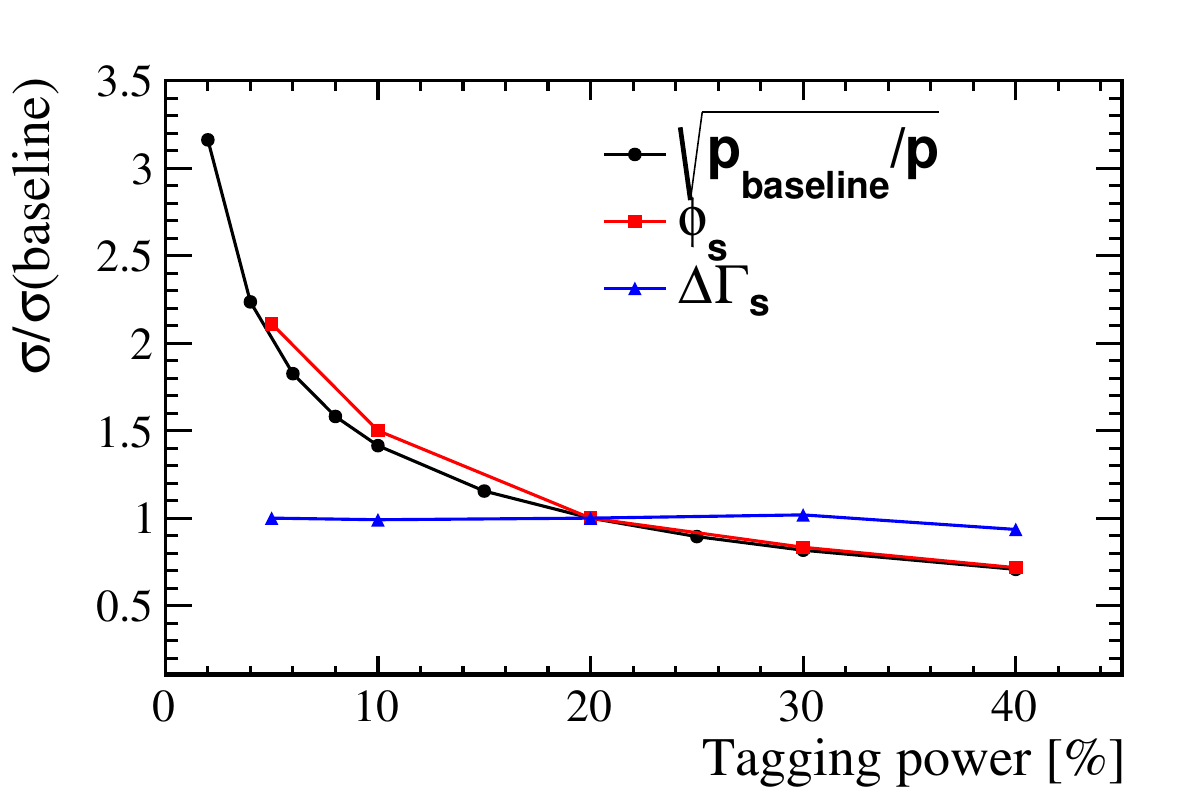}
\end{center}
\caption{
The varying precision for $\phis$ and $\Delta\Gamma_s$ as a function of the decay time resolution (left) and tagging power (right). The ratio to the baseline precision is plotted. The baseline precision is with the parameters $\sigma_t = 4.7~\mathrm{fs}$ and $p = 20\%$. The red line with a square marker and the blue line with a triangle marker represent the precision from the toy Monte Carlo simulation, respectively. The black line with a circle marker represents the precision from the analytical formula.}
\label{fig:res}
\end{figure}

The dependent of $\sigma(\phis)$, $\sigma(\Delta\Gamma_s)$ and $\sigma(\Gamma_s)$ on the time resolution and tagging power is investigated with toy Monte Carlo simulation.
Figure~\ref{fig:res} shows the varying resolution for \phis and $\Gamma_s$ as a function of the tagging power and decay time resolution. The ratio to the baseline resolution is plotted. The baseline resolution is with the parameters 
$\sigma_t = 4.7~\mathrm{fs}$ and $p = 20\%$. The red line with a square marker and the blue line with a triangle marker represents the resolution from the toy Monte Carlo simulation, respectively. The black line with a circle marker represents the resolution from the analytical formula. 
The resolution ratio of $\Gamma_s$ is almost the same as that of $\Delta\Gamma_s$.

The simulation provides a validation of the formula
\begin{displaymath}
    \sigma(\phis) \propto 
    1/\exp({-\frac12\Delta m_{s}^2\sigma_t^2})
\end{displaymath}
and 
\begin{displaymath}
    \sigma(\phis) \propto 
    1/\sqrt{p},
\end{displaymath}
and it also provides a validation that the precision of $\Gamma_s$ and $\Delta\Gamma_s$ are insensitive to the time resolution and tagging power. 

\begin{figure}[!hbt]
    \begin{center}
\includegraphics[width=0.5\textwidth]{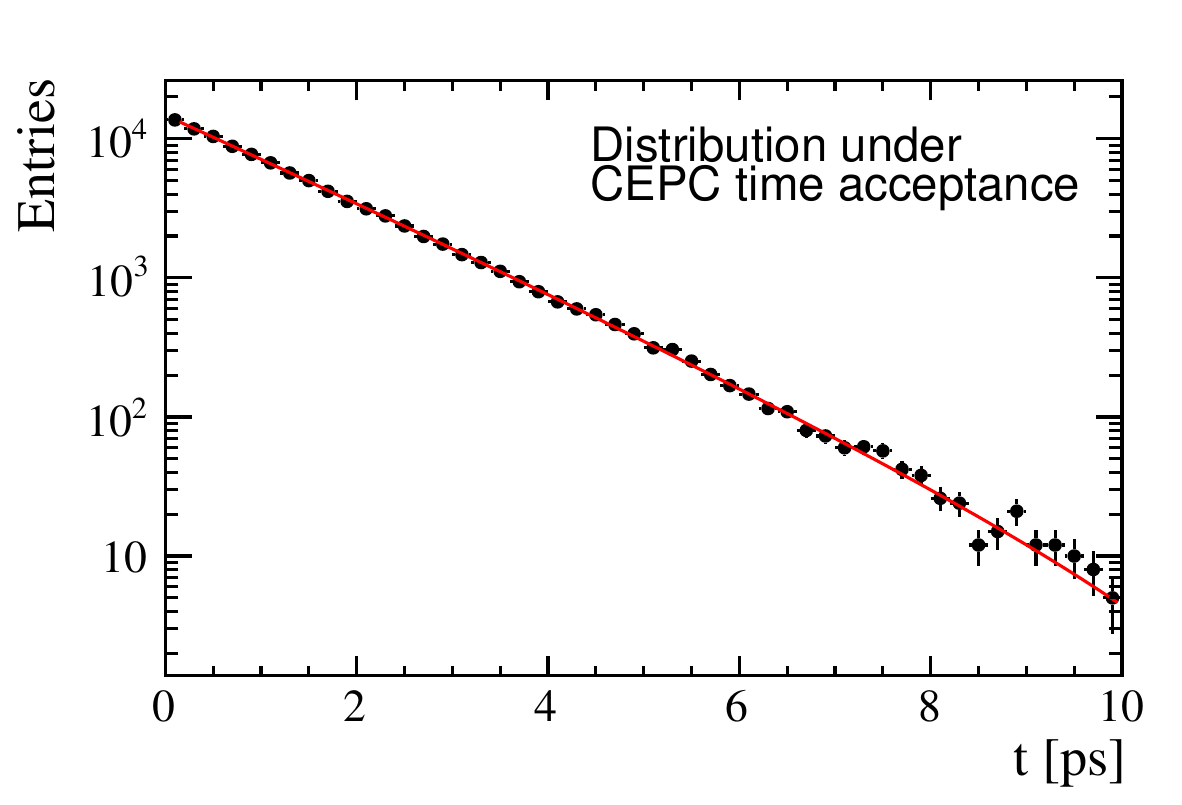}
\includegraphics[width=0.5\textwidth]{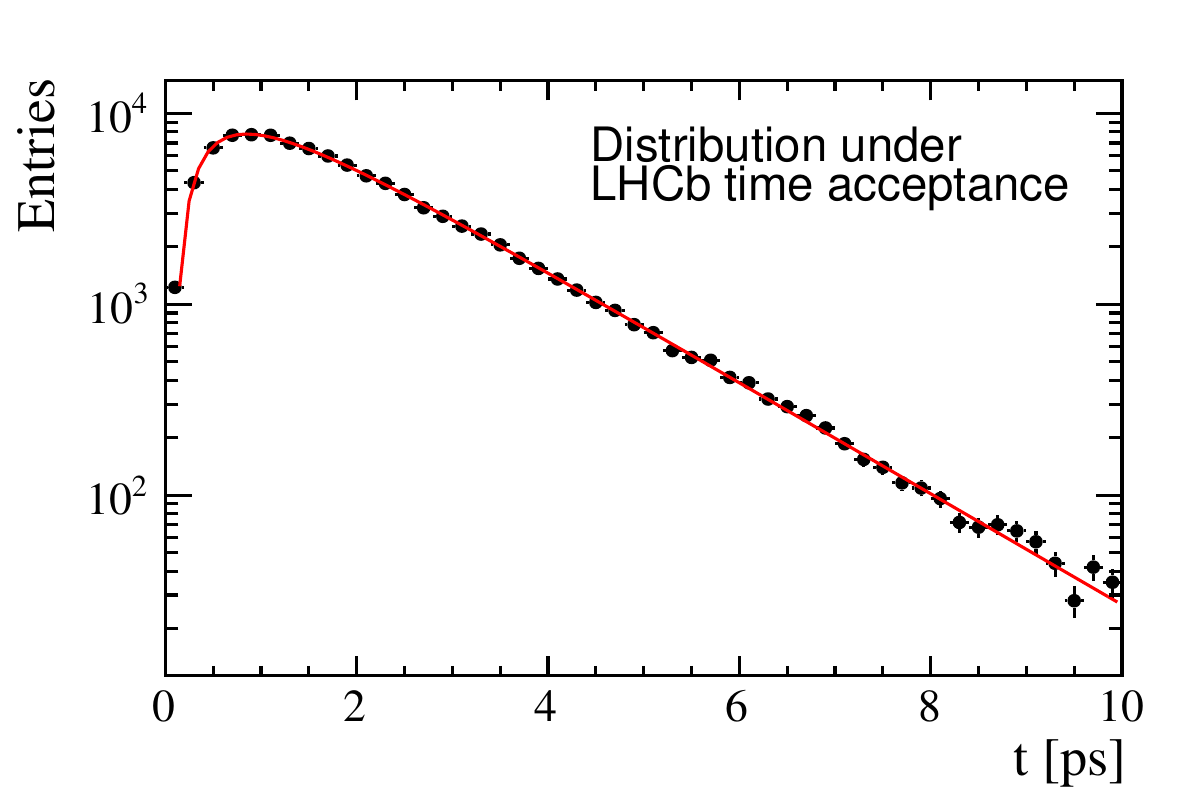}
\end{center}
\caption{Events generated with the distribution $f_\mathrm{acc} \exp(-t/\tau)$ where $f_\mathrm{acc} = f_\mathrm{acc}^\mathrm{CEPC}$ (top) and $f_\mathrm{acc} = f_\mathrm{acc}^\mathrm{LHCb}$ (bottom). The events are fitted to the respective model $f_\mathrm{acc} \exp(-t/\tau)$.}
\label{fig:fit}
\end{figure}

The influence of decay time acceptance on the precision of $\Gamma_s$ is examined. Two samples, each consisting of $10^5$ events, are generated with the distributions $f_\mathrm{acc}^\mathrm{CEPC} \exp(-t/\tau)$ and $f_\mathrm{acc}^\mathrm{LHCb} \exp(-t/\tau)$, respectively. The parameter $\tau$ is set as $\tau=1.538~\mathrm{ps}$. These events are then fitted to the models $f_\mathrm{acc}^\mathrm{CEPC} \exp(-t/\tau)$ and $f_\mathrm{acc}^\mathrm{LHCb} \exp(-t/\tau)$ with $\tau$ as a free parameter, as shown in Fig.~\ref{fig:fit}.
The study yielded a $\sigma(\tau) = 0.0058~\mathrm{ps}$ using the CEPC time acceptance profile, and $\sigma(\tau) = 0.0049~\mathrm{ps}$ for the LHCb time acceptance profile.

\section*{Acknowledgements}
We would like to thank Jibo He, Wenbin Qian, Yuehong Xie, and Liming Zhang for their help in the discussion, polishing the manuscript, and cross-checking the results.

% BibTeX users please use
% \bibliographystyle{unsrt}
\bibliographystyle{sn-nature}
\bibliography{main}

\end{document}